\documentstyle[prd,preprint,tighten,aps,aps10,aps12,eqsecnum,amssymb,%
newlfont,prabib,epsfig,graphicx]{revtex}

\newcommand{\beq}{\begin{equation}}
\newcommand{\eeq}{\end{equation}}
\newcommand{\beqa}{\begin{eqnarray}}
\newcommand{\eeqa}{\end{eqnarray}}
\newcommand{\beqan}{\begin{eqnarray*}}
\newcommand{\eeqan}{\end{eqnarray*}}
\newcommand{\ba}{\begin{array}}
\newcommand{\ea}{\end{array}}
\newcommand{\no}{\nonumber}
\newcommand{\ra}{\rightarrow}
\newcommand{\Ra}{\Rightarrow}
\newcommand{\ve}{\varepsilon}
\newcommand{\A}{{\cal A}}
\newcommand{\B}{{\cal B}}
\newcommand{\Ha}{{\cal H}}
\newcommand{\cO}{{\cal O}}

\normalsize
\begin{document}
\preprint{
\begin{tabular}{r}
UWThPh-2001-47 \\
November 2001
\end{tabular}
}
\draft
\title{A Geometric Picture of Entanglement \\ and Bell Inequalities}
\author{R.A. Bertlmann, H. Narnhofer and W. Thirring}
\address{Institut f\"ur Theoretische Physik \\ Universit\"at Wien\\
Boltzmanngasse 5, A-1090 Wien}

\maketitle

\vfill

\begin{abstract}
We work in the real Hilbert space $\Ha_s$ of hermitian Hilbert-Schmidt operators and show
that the entanglement witness which shows the maximal violation of a generalized Bell
inequality (GBI) is a tangent functional to the convex set $S\subset \Ha_s$ of separable
states. This violation equals the euclidean distance in $\Ha_s$ of the entangled state to
$S$ and thus entanglement, GBI and tangent functional are only different aspects of the
same geometric picture. This is explicitly illustrated in the example of two spins,
where also a comparison with familiar Bell inequalities is presented.
\end{abstract}

\vspace{1cm}


\vfill

\section{Introduction}

The importance of entanglement \cite{Schrodinger,EPR} of quantum states became quite
evident in the last ten years. It is the basis for such physics, like quantum cryptography
\cite{Ekert,DeutschEkert,Hughes,GisinGroup} and quantum teleportation
\cite{Bennett,ZeilingerTele}, and it triggered a new technology:
quantum information \cite{ZeilingerInfo1,ZeilingerInfo2}.
Entangled states lead to a violation of Bell inequalities (BI) which distinguish
quantum mechanics from (all) local realistic theories \cite{Bell2}.
Much effort has been made in studying the mathematical structure of entanglement,
especially the quantification of entanglement (see, for instance, Refs.
\cite{HHHbook,Terhal}).
There exist different kinds of measures of entanglement indicating somehow the
difference between entangled and separable states, which is usually related to the
entropy of the states (see, e.g.,
Refs.\cite{Bennett1,Holevo,VedralKnight,Rudolph,Narnhofer,HHH99}).
In this paper we define a simple and quite natural measure for entanglement,
a distance of certain vectors in Hilbert space which has as elements both observables
and states, and we relate it to the maximum violation of a generalized Bell inequality
(GBI). We work with a bipartite system in a finite-dimensional Hilbert space but
generalizations are possible.

The Hilbert-Schmidt distance $D$ of a state to the set of separable states has previously
been proposed as a measure of entanglement \cite{WitteTrucks,Ozawa}. Our point is that
if one admits all of $\B (\Ha_A \otimes \Ha_B)$ as entanglement witnesses then the
maximal violation $B$ of the corresponding GBI equals the distance $D$ numerically.
Since $D$ can be written as a minimum and $B$ as a maximum upper and lower bounds
are readily available. In fact, in some standard examples one can make them coincide
and thus calculate $B = D$ exactly.

Though distinct from the entropic entanglement descriptions, the Hilbert-Schmidt distance
$D$ as a quantitative description of entanglement is insofar reasonable, as considered
as functional of the state it is convex and invariant under local unitary transformations.
This implies that states more mixed in the sense of Uhlmann \cite{Uhlmann} have a lower
entanglement. However, $D$ is not monotonic decreasing under arbitrary completely positive
maps in $\Ha_A$ or $\Ha_B$ but only if they have norm one. Thus whether they satisfy
monotonicity in ``local operations and classical communication'' depends on the exact
definition of this term.\\

We consider a finite-dimensional Hilbert space $\Ha = {\mathbf C}^{N}$, where
observables $A$ are represented by all Hermitian matrices and states $w$ by density
matrices. It is useful to regard these quantities as elements of a real
Hilbert space $\Ha_s = {\mathbf R}^{N^2}$ with scalar product
\beq
( w|A ) = \mbox{Tr } w A
\eeq
and corresponding norm
\beq
\| A \|_2 = (\mbox{Tr }A^2)^{1/2}
\eeq
(we identify quantities with their representatives in $\Ha$). Both density matrices
and observables are represented by vectors in $\Ha_s$, a density matrix is positive
and has trace unity.

Unitary operators $U$ in $\Ha$ induce via $U A U^\ast = O A$ orthogonal operators
$O$ in $\Ha_s$, but the homomorphism $U \to O$ is neither injective nor surjective.

\section{Spin examples}

Let us begin with two examples which will be of our interest.\\

\noindent
{\bf Example I:} One spin.

\noindent
Generally an observable can be written as
\beq
A = \alpha \, {\mathbf 1} + \vec a \cdot \vec \sigma \, , \quad \mbox{with} \quad \alpha \in
{\mathbf R} \, , \, \vec a \in {\mathbf R}^3.
\eeq
The operator $A$ is a density matrix iff
$\alpha = 1/2$ and $\|\vec a\| \leq 1/2$, it gives a pure state iff
$\| \vec a \| = 1/2$ or $A^2 = A$.
If the state is
\beq
w = \frac{1}{2} \, ({\mathbf 1} + \vec w \cdot \vec \sigma)
\eeq
the expectation value of $A$ is
\beq
( w|A ) = \alpha + \vec a \cdot \vec w \, .\\
\eeq

For us the important structural element is a tensor product
$\Ha = \Ha_A \otimes \Ha_B$ which defines the set $S$ of separable
(classically correlated) states $\rho_A^i \, ,\rho_B^j$
\beq
S = \Big\{ \rho = \sum_{i,j} c_{ij} \, \rho_A^i \otimes \rho_B^j \, \Big\vert \,
0 \leq c_{ij} \leq 1, \sum_{i,j} c_{ij} = 1\Big\}.\\
\eeq

\noindent
{\bf Example II:} Two spins $\vec \sigma_A$ and $\vec \sigma_B \,$,
``Alice and Bob''.

\noindent
An observable $A$ can be represented by
\beq\label{observable}
A = \alpha \, {\mathbf 1} + a_i \, \sigma_A^i \otimes {\mathbf 1}_B
+ b_i \, {\mathbf 1}_A \otimes \sigma_B^i + c_{ij} \, \sigma_A^i \otimes \sigma_B^j \, ,
\eeq
\beq
\frac{1}{4} \,  \| A \|_2^2 = \alpha^2 + \sum_i (a_i^2 + b_i^2) + \sum_{i,j} c_{ij}^2 \, .
\eeq
Note that $c_{ij}$ can be diagonalized by $2$ independent orthogonal
transformations on $\sigma_A^i$ and $\sigma_B^j$ \cite{HenleyThirring}.
The operator $A$ is a density matrix if $\alpha = 1/4$ and the operator norm
$\|\;\|_\infty$ of $A - 1/4$ is $\leq 1/4$.
Since $\|\;\|_2 \geq \|\;\|_\infty$ this is satisfied if
\beq
\sum_i (a_i^2 + b_i^2) + \sum_{i,j} c_{ij}^2 \leq 1/16 \, .
\eeq
For pure states $\| \; \|_2 = \| \; \|_\infty$ and $\|\rho\|_2 = 1$ is
necessary and sufficient for purity. A pure separable state has the form
\beq
\rho = \frac{1}{4} \, \big({\mathbf 1} + n_i \, \sigma_A^i \otimes {\mathbf 1}_B
+ m_i \, {\mathbf 1}_A \otimes \sigma_B^i  + n_i m_j \,
\sigma_A^i \otimes \sigma_B^j\big) \, ,
\eeq
with $\vec n^2 = \vec m^2 = 1$, and gives the expectation value of $A$
\beq
( \rho|A ) = \alpha + \vec n \cdot \vec a + \vec m \cdot \vec b + n_i m_j \,
c_{ij} \, .
\eeq

\section{Generalized Bell inequality}

States which are not separable are called entangled $w \in S^c$,
the complement in the set of states.
We introduce as a measure of entanglement $D(w)$ the $\Ha_s$-distance of $w$
to the set $S$ of separable states
\beq\label{distance}
D(w) = \min_{\rho \,\in S} \|\, \rho - w \, \|_2 \, .
\eeq
Since
$$
\|\, \rho - w \, \|_2^2 \, =\, \mbox{Tr }\big( \,\rho^2 + \omega^2 - 2 \, \sqrt{\rho} \;
\omega \, \sqrt{\rho} \, \big) \, \leq \, \mbox{Tr }\big( \rho^2 + \omega^2 \big) \leq 2
$$
we generally have
\beq
0 \leq D(w) \leq \sqrt{2} \, .
\eeq
Usually the Bell inequality refers to an operator in the tensor product where
by classical arguments only some range of expectation values can be expected
whereas quantum mechanics permits other values.
A Bell inequality in a generalized sense is given by an observable $A \not\geq 0$
for which
\beq\label{BI}
( \rho|A ) \geq 0 \quad \forall \rho \in S \, .
\eeq
Thus $\exists \, w$ such that
\beq \label{BIviol}
( w|A ) < 0 \quad \mbox{for some} \quad w \in S^c \, .
\eeq
Such elements $A \in \A_W$ are called entanglement witnesses \cite{HHH96,TerhalPhysLett}.
A product operator can never be $\in \A_W$ but already the sum of
two products serves for the CHSH (Clauser, Horne, Shimony, Holt) inequality 
\cite{CHSH}. But the number of summands is not restricted in $\A_W$.
The operator $A \in \A_t$ becomes a tangent functional if in addition
$\exists \, \rho_0 \, \in S$ such that $( \rho_0|A ) = 0$.
Since $S$ is a convex subset of the state space such tangential $A$'s always exist.
Even more, the set $S$ is characterized by the tangent functionals and the
$\rho_0$'s with $( \rho_0|A ) = 0$, for some $A \in \A_t \, ,$ are the boundary
$\partial S$ of $S$.

Frequently a bigger set than $S$ is considered as classically explainable
in a local hidden variable theory. Bell inequalities are those which
contradict even those sets. To avoid misunderstandings we call generalized
Bell inequalities expectation values which contradict the predictions from
$S$, the set of separable states.

Thus the GBI (\ref{BI}) is violated by an entangled state $w$,
Eq.(\ref{BIviol}), and we get the following inequality for some $A \in \A_W$
\beq\label{GBI}
( \rho|A ) > ( w|A )  \quad \forall \rho \in S \, .
\eeq
Considering now the maximal violation of the GBI
\beq\label{BImax}
B(w) = \max_{\|A - \alpha \|_2 \leq 1} \,
\lbrack \, \min_{\rho \,\in S} ( \rho|A ) - ( w|A ) \, \rbrack \, ,
\eeq
we find the following result\\

\noindent
{\bf Theorem:}

i) The maximal violation of the GBI is equal to the distance of $w$ to the set $S$
\beq\label{Theorem}
B(w) = D(w)  \quad \forall \; w \ .
\eeq

ii) The min of $D$ is attained for some $\rho_0$ and the max of $B$ for
\beq\label{Amax}
A_{max} = \frac{\rho_0 - w - ( \rho_0|\rho_0 - w ){\mathbf 1}}{\| \rho_0 - w \|_2}
\in \A_t \, .
\eeq

iii) For $B = D$ the following two-sided variational principle holds
\beq\label{TheoremVar}
\min_{\rho \,\in S} \left( \rho - w \, \bigg| \, \frac{\rho^{\, '} - w}%
{\| \rho^{\, '} - w \|_2} \right)
\leq B(w) \leq \| \rho^{\, '} - w \|_2 \qquad \forall \rho^{\, '} \in S \, .
\eeq
(For an illustration, see Fig. 1; for a similar view, particularly about $A_{max}$,
see Ref. \cite{PittengerRubin}).\\

\noindent
{\bf Remark:}

\noindent
The proof of the Theorem does not use the product structure of the Hilbert space
$\Ha$ but only the geometric properties of the Euclidean distance in $\Ha_s$.
It can be illustrated already with one spin where the set of separable states $S$ is
replaced by $S_z$
\beq
S_z = \Big\{ \rho = \frac{1}{2} ({\mathbf 1} + \lambda \sigma_z) , \,
|\lambda| \leq 1\Big\}\, ,
\eeq
and
\beq
w = \frac{1}{2} \, ({\mathbf 1} + \vec w \cdot \vec \sigma) \, , \qquad \|w\|_2 \leq 1
\eeq
is considered as the analogue of an entangled state, if $w_x$ or $w_y \ne 0$.

\noindent
The observables $A$ with $\|A\|_2 = 1$ are of the form
\beq
A = \frac{\alpha {\mathbf 1} + \vec a \cdot \vec \sigma}{\sqrt{2} \, (\alpha^2 + a^2)^{1/2}} \, ,
\quad \mbox{and} \quad a = \| \vec a \| \, , \, \vec a \in {\mathbf R}^3.
\eeq
For the $\Ha_s$-distance $D$, our measure of entanglement, we calculate
$$
\min_{\rho} \|\rho - w\|_2^2 =
\min_\lambda \frac{1}{4} \|\lambda \sigma_z  - \vec w \cdot \vec \sigma\|_2^2  =
\min_\lambda \frac{1}{2} \big( (\lambda - w_z)^2 + w_x^2 + w_y^2 \big) =
\frac{1}{2} (w_x^2 + w_y^2)
$$
attained for $\lambda = w_z$, so that we have
\beq\label{distance1spin}
D(w) = \frac{1}{\sqrt{2}} \, (w_x^2 + w_y^2)^{1/2} \, .
\eeq

\begin{figure}
\center{
\includegraphics[width=10cm,height=7cm,keepaspectratio=true]{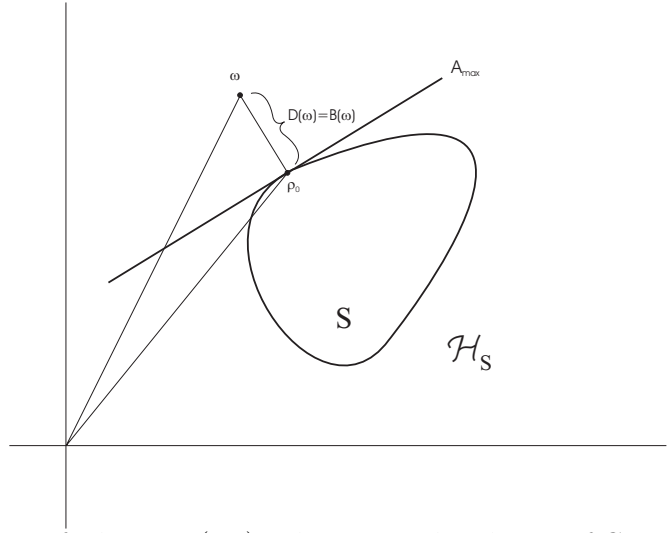}
}
\caption{Illustration of Theorem (\ref{Theorem}). The maximal violation of
GBI $B(w)$, Eq.(\ref{BImax}), which is equal to the $\Ha_s$-distance $D(w)$,
Eq.(\ref{distance}), of an entangled state $w$ to the set $S$ of separable states,
is shown together with the tangent plane defined by $A_{max}$ (\ref{Amax}).}
\end{figure}

\vspace{1.5cm}

On the other hand, we find for the maximal violation of the GBI
\beqa\label{BI1spin}
B(w) &=& \max_{\vec a, \, \alpha} \min_\lambda \frac{1}{2} \left(\lambda \sigma_z -
\vec w \cdot \vec \sigma \, \bigg| \,
\frac{\alpha {\mathbf 1} + \vec a \cdot \vec \sigma}{\sqrt{2}(\alpha^2 + a^2)^{1/2}} \right) =
\max_{\vec a, \, \alpha} \frac{-1}{\sqrt{2}} \cdot \frac{|a_z| + \vec w \cdot \vec a}
{(\alpha^2 + a^2)^{1/2}} \no \\
&=& \frac{1}{\sqrt{2}} \, (w_x^2 + w_y^2)^{1/2} \, .
\eeqa
Here the observable
\beq
A_{max} = - \frac{w_x \, \sigma_x + w_y \, \sigma_y}{\sqrt{2} \,(w_x^2 + w_y^2)^{1/2}}
\eeq
is the tangent functional $\forall \rho \in S_z$, $\partial S_z = S_z$.

Note that for the maximal violation of the GBI (\ref{BI1spin}) the $\min_{\rho \in S}$
is attained for $\frac{1}{2}(1 - \sigma_z)$ if $a_z > 0$ and not for
$\frac{1}{2}(1 + w_z \sigma_z)$ as
in case of the distance (\ref{distance1spin}). It means that for $D$ the
$\min_{\rho}$ is not necessarily attained for a pure state but for $B$ it is
since it is effectively a max. Thus the equality $B = D$, Theorem part (\ref{Theorem}), is
not so trivial since the extrema may be attained at disjointed sets. Then min max may be
bigger than max min as can be seen already in $\min_{i}$ and $\max_{j}$ for the matrix
$$
M_{ij} = \left( \begin{array}{cc}
0 & 1 \\
1 & 0
\end{array} \right) \, .\\
$$

\noindent
{\bf Proof of the Theorem:} Eq.(\ref{Theorem})

\noindent
$D(w) = \min_{\rho \in S} \| \rho - w \|_2$ is attained for some $\rho_0$ since
$\|\;\|_2$ is continuous and $S$ is compact. Now take for
$A - \alpha = ( \rho_0 - w ) / \| \rho_0 - w \|_2$ in the definition of $B$ and use the
orthogonal decomposition with respect to this unit vector,
$\Ha_s \ni v = v_{\|} + v_\perp$, $( v_\perp | \rho_0 - w ) = 0$.
Therefore we can apply simple Euclidean geometry and decompose the vector
$\rho - w$ in the above sense.

We also remember that $\rho_0 -w$ is the normal to the tangent plane to $S$,
which means
$$
\|(\rho - w)_\| \|_2 \ge \|(\rho_0 - w)_\| \|_2 = \| \rho_0 - w \|_2
$$
since $S$ is convex, see Fig. 2. This we can prove in the following way.
The tangent $A_{max}$ divides the state space into
$\Ha_w = \{\rho : \|(\rho - w)_\| \|_2 < \|(\rho_0 - w)_\| \|_2 \}$, which contains $w$,
and $\Ha_w^c$, the complement to $\Ha_w$. If $\Ha_w$ were to contain $\rho \in S$
then because of the convexity of $S$ it would contain all
$\rho_{\lambda} = (1 - \lambda) \rho_0 + \lambda \rho , \quad
\lambda \in \lbrack 0,1 \rbrack$.
Since $\rho_{\lambda}$ would have an angle of less than $90^0$ with $\rho_0 -w$
there would be a $\rho_{\lambda}$ inside the ball
$\|(\rho - w) \|_2 < \|(\rho_0 - w) \|_2 = D(w)$ and $\rho_0$ would not be the point
of $S$ of minimal distance to $w$. Therefore $S \subset \Ha_w^c$ and
$\|(\rho - w)_\| \|_2 \ge \|(\rho_0 - w) \|_2 \quad \forall \rho \in S$.

Using above arguments we obtain
$$
B(w) \geq \min_\rho \left( \rho - w \, \bigg| \,
\frac{\rho_0 - w}{\| \rho_0 -w \|_2} \right) \geq
\min_\rho  \left(  (\rho - w)_\| \, \bigg| \,
\frac{\rho_0 - w}{\| \rho_0 -w \|_2} \right) \\
$$
$$
\geq  \left( \rho_0 - w \, \bigg| \,
\frac{\rho_0 - w}{\| \rho_0 -w \|_2} \right) = \| \rho_0 - w \|_2 = D(w) \, .
$$
On the other hand, $D$ and $B$ can be written as $\min_\rho \max_A$ and
$\max_A \min_\rho$ of $( \rho - w | A )$ and generally we have
$\min \max \geq \max \min$ . So a priori we know $D(w) \geq B(w)$ and
we conclude $D(w) = B(w)$.
\begin{flushright}
$\Box$
\end{flushright}

\section{Properties of the generalized Bell inequality}

Now we discuss the properties of $D(w)$, Eq.(\ref{distance}), the $\Ha_s$-distance
of $w$ to the set $S$ of separable states, which is equal to $B(w)$, Eq.(\ref{BImax}),
the maximal violation of the GBI.\\

\noindent
{\bf Properties of $D(w)$:}
\begin{enumerate}
\item[i)] $D(w)$ is convex,
\item[ii)] $D(w)$ is continuous,
\item[iii)] $D(w) = D(U_A \otimes U_B \, w \, U_A^* \otimes U_B^*)$ $\quad \forall$
unitary operators $U_{A,B} \,$.
\item[iv)] $D(w)$ is monotonic decreasing under mixing enhancing maps, see e.g. Ref.
\cite{HaydenTerhalUhlmann}.\\
\end{enumerate}

\begin{figure}
\center{
\includegraphics[width=10cm,height=7cm,keepaspectratio=true]{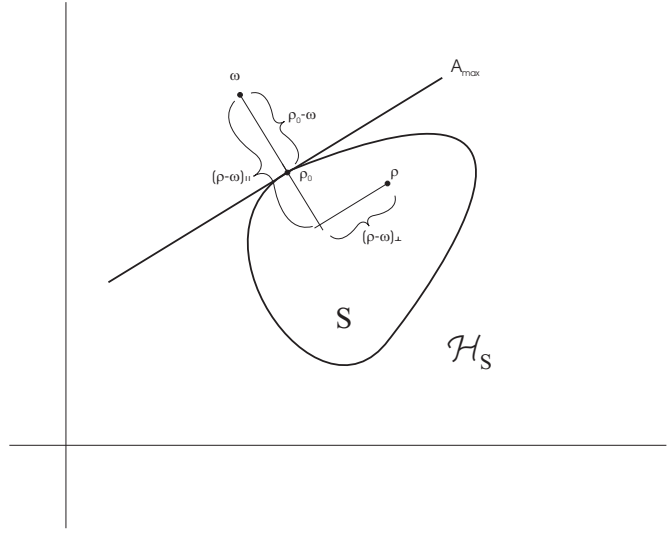}
}
\caption{For illustration we have drawn the vectors used in the
Proof of Theorem (\ref{Theorem}).}
\end{figure}

\noindent
{\bf Remarks:}
\begin{enumerate}
\item[ad i)] It means that by mixing the entanglement decreases and the
maximally entangled states must be pure. This is to be expected since the
tracial state $w_{\mathrm{tr}} = 1/(\dim \Ha_A \dim \Ha_B)$ is separable
$\Leftrightarrow D(w) = 0$.\\
Furthermore the set $\{ w \, |\, D(w) < c \}$ is convex.
\item[ad ii)] It tells us for an entangled state a neighborhood of it
is also entangled. Actually a neighborhood of the tracial state is also
separable.
\item[ad iii)] The state space decomposes into equivalence classes of
states with the same entanglement. All pure separable states are in the
same equivalence class.
\item[ad iv)] Mixing enhancing maps are essentially a combination of unitary transformations
and convex combinations.\\
\end{enumerate}

\noindent
{\bf Proofs of the B, D properties:}
\begin{enumerate}
\item[i)] $B(w)$ and $D(w)$ are continuous:
$$
| (w + \delta - \rho | A ) - ( w - \rho | A ) | \leq \ve \quad \forall \;
\|\delta\|_2 \leq \ve, \; \| A \|_2 \leq 1
$$
$$
\Ra \quad |(B \, \mbox{or} \, D)(w + \delta) - (B \, \mbox{or} \, D )(w)| \leq \ve
\quad \forall \; \|\delta \|_2 \leq \ve \, .
$$
\item[ii)] $B(w)$ is convex:
$$
B \Big( \sum_i \lambda_i w_i \Big) = \max_A \sum_i \lambda_i
\left( \min_{\rho \in S} ( \rho |A ) - ( w_i | A ) \right)
$$
$$
\leq\sum_i \lambda_i \left(\max_A \,
\lbrack \, \min_{\rho \in S} ( \rho |A ) - ( w_i | A ) \, \rbrack \right)
= \sum_i \lambda_i \; B(w_i) \, .
$$
\item[iii)] $D(w) = D(U_A \otimes U_B \, w \, U_A^* \otimes U_B^*)$
follows from the invariance of $S$ under $U_A \otimes U_B \,$.
\item[iv)] The monotonic decrease under mixing enhancing maps is a consequence of
points i) and iii).
\end{enumerate}
\begin{flushright}
$\Box$
\end{flushright}

The ``most'' separable state $w_{\mathrm{tr}} = 1 / dim \Ha$ is a convex
combination of most entangled states. From the properties of $D(w)$ we get
the following artistic impression. In the state space there is a plane around
$w_{\mathrm{tr}}$ with $D(w) = 0$. From it emerge valleys with $D(w) = 0$ to
the pure separable states on the boundary. In their neighborhood
are entangled states thus $D$ slopes up in such a way
that the regions $D \leq c$, with $0 \leq c \leq D_{\mathrm{max}}$,
are convex. On the boundary of the state space also sit the states with
$D = D_{\mathrm{max}}$ and they form a rim. Since $U_A \otimes U_B$ act
continuously in a neighborhood of maximally entangled states there are
others with $D = D_{\mathrm{max}}$ but also some with $D < D_{\mathrm{max}}$ which
one gets by mixing in a little bit with the separable states.\\

This somewhat poetic description is mathematically supplemented by considering
$S$ as a subset of the state space $S \cup S^c \subset \Ha_s$, so the boundary
$\partial S$ are those elements of $S$ where in each neighborhood there are
entangled states.

\section{Geometry of separable states}

What is the geometric structure of the set $S$ of separable states?
Let us investigate its properties.\\

\noindent
{\bf Properties of S:}
\begin{enumerate}
\item[i)] The dimensions of both $S$ and $S^c$ are $N^2 -1$.
\item[ii)] Pure separable states belong to the boundary $\partial S$ and convex
combinations of two of them are still on $\partial S$.
\item[iii)] If a mixture $\rho = \sum_{i=1}^{n} \mu_i \rho_i$ is on $\partial S$
then there is a face, i.e.
$$
\bar\rho = \sum_{i=1}^{n} \bar\mu_i \rho_i
\in \partial S \quad \forall \; \bar\mu_i \geq 0, \,
\sum_{i=1}^{n} \bar\mu_i = 1 \, .
$$
\item[iv)] If $\Ha_A = \Ha_B$ ($= {\mathbf C}^{\sqrt{N}}$) then $\partial S$ contains
at least $N$ dimensional faces.
\item[v)] $S$ is invariant under $T_A \otimes {\mathbf 1}_B$ , with $T_A$
any positive map $\B (\Ha_A) \to \B (\Ha_A)$ .
\item[vi)] If $A \not\geq 0$ but $(T_A \otimes {\mathbf 1}_B) A \ge 0$ then
$A \in \A_W$ and if $\exists \, \rho_0 \, \in S$ such that $( \rho_0|A ) = 0$
then $A \in \A_t$ .\\
\end{enumerate}

\noindent
{\bf Remarks:}
\begin{enumerate}
\item[ad i)] It means that both $S$ and $S^c$ are everywhere thick and
do not have pieces of lower dimensions.
\item[ad ii)] Clearly the convex combination of two pure states lies
(for $N > 2$) on the boundary of the state space since in each neighborhood
there are not positive functionals. Here we have the stronger statement that
in each neighborhood there are entangled states.
\item[ad iii)] If $\partial S$ has a n-dimensional flat part this means that
mixtures of n pure states are on $\partial S$. Point iii) affirms the converse
in the sense that in the decomposition the $\rho_i$'s span a face.
\item[ad iv)] It says that $n = N$ actually occurs.
\item[ad v)] Strangely, the tensor product of two positive maps is not necessarily
positive but applied to separable states it is.\\
\end{enumerate}

\noindent
{\bf Proofs of the properties of $S$:}
\begin{enumerate}
\item[i)] $S$ has the full dimension of $N$ since a neighborhood of the tracial
state $w_{\mathrm{tr}} = 1/N$ is separable and as a convex set it has everywhere the same
dimension. The complement $S^c$, the set of entangled states, has the full dimension
since $D$ is continuous and if $D(w) > 0$ it is so for a neighborhood of $w$.
\item[ii)] $\rho$ is pure and separable $\, \Ra \, \rho \in \partial S \, $:\\
If $\rho = | \phi \otimes \psi \rangle \langle \phi \otimes \psi |$ (pure and
separable) then $| \phi \otimes \psi + \ve \phi^{'} \otimes \psi^{'} \rangle
\langle \phi \otimes \psi + \ve \phi^{'} \otimes \psi^{'} |$ comes for $\ve \to 0$
arbitrarily close and is $\forall \, \ve$ pure and not a product state
$\, \Ra$ it is entangled $\, \Ra \, \rho \in \partial S$.\\
$\rho_{\, i}$ is pure and separable $\, \Ra \, \rho_{\lambda} = \lambda \rho_1 +
(1 - \lambda) \rho_2 \in \partial S \, $:\\
Let us take
$\rho_{\, i} = | \phi_i \otimes \psi_i \rangle \langle \phi_i \otimes \psi_i |$
and consider
$$
\lambda | \phi_1 \otimes \psi_1 + \ve \phi_2 \otimes \psi_2 \rangle
\langle \phi_1 \otimes \psi_1 + \ve \phi_2 \otimes \psi_2 |
+ (1 - \lambda) | \phi_2 \otimes \psi_2 + \ve^{'} \phi_1 \otimes \psi_1 \rangle
\langle \phi_2 \otimes \psi_2 + \ve{'} \phi_1 \otimes \psi_1 | \, .
$$
For $\ve, \, \ve^{'} \to 0$ it comes arbitrarily close to $\rho_{\lambda}$
but in the two dimensional Hilbert subspace spanned by $\phi_i \otimes \psi_i$
($i = 1,2$) the only separable pure states are of the form $\rho_{1,2} \,$.
Thus a state that is not a linear combination of $\rho_1$ and $\rho_2$
needs for its decomposition into pure states at least one pure entangled state,
and is therefore entangled itself.
Therefore we have an entangled state arbitrarily close to $\rho_{\lambda} \,$
$\, \Ra \, \rho_{\lambda} \in \partial S \, $
(compare with Refs.\cite{HillWootters,BNU}).
\item[iii)] For a tangent functional $A$ at $\rho = \sum \mu_{\, i} \rho_{\, i} \,$,
$\rho_{\, i} \in S$, we have
$$
0 = (\rho | A ) = \sum \mu_{\, i} (\rho_{\, i} | A ) \quad \Ra \quad
(\rho_{\, i} | A )= 0 \quad \forall \, i
$$
$$
\Ra \quad \left( \sum \bar \mu_{\, i} \rho_{\, i} \Big| A \right) = 0 \quad \Ra \quad
\sum \bar \mu_{\, i} \rho_{\, i} \in \partial S \, .
$$
\item[iv)] For a given tangent functional $A_t = A_1 - A_2 \, , A_i \ge 0 \, ,
\|A_2\|_2 = 1$ there exists an entangled state $w$ with $(w|A_t) = - (w|A_2)
\le -1 + \ve \,$. The homotopic state $\bar w = (1 - \frac{\ve}{2})w +
\frac{\ve}{2}w_{\mathrm{tr}}$ is also entangled
since $D(\bar w)$ is continuous, and the corresponding density matrix is invertible
and needs $N$ components to be decomposed into pure states. There exists a
continuous path from the entangled $\bar w$ to the separable $w_{\mathrm{tr}}$ formed from
states with corresponding invertible density matrices. When this path passes the
boundary $\partial S$ then according to property iii) we obtain a separable state
embedded in a N-dimensional face of $\partial S$.
\item[v)] Follows from the results in Ref.\cite{HHH96}.
\item[vi)] Follows from v) and the definitions of $\A_W$ and $\A_t \,$.
\end{enumerate}
\begin{flushright}
$\Box$
\end{flushright}

\section{Geometry of entangled and separable states of spin systems}

We focus again on the two spin example and calculate the entanglement of the
following quantum states.\\

\noindent
{\bf Example:} Alice and Bob, the ``Werner states''.

\noindent
Let us consider Werner states \cite{Werner} which can be parameterized by
\beq\label{Wstate}
w_\alpha = \frac{{\mathbf 1} - \alpha \, \vec \sigma_A \otimes \vec \sigma_B}{4} \, ,
\eeq
and they are possible density matrices for $-1/3 \leq \alpha \leq 1$ since
$\vec \sigma_A \otimes \vec \sigma_B$ has the eigenvalues $-3,1,1,1$.
To calculate the entanglement we first mix product states to get
$$
\frac{1}{2} \left\{ \frac{({\mathbf 1}_A - \sigma_A^x) \otimes ({\mathbf 1}_B + \sigma_B^x)}{2}
+ \frac{({\mathbf 1}_A + \sigma_A^x) \otimes ({\mathbf 1}_B - \sigma_B^x)}{2} \right\} =
\frac{{\mathbf 1} - \sigma_A^x \otimes \sigma_B^x}{4}
$$
and then with $x \ra y$, $x \ra z$ finally
\beq\label{rho0}
\rho_0 = \frac{1}{4} \, ({\mathbf 1} - \frac{1}{3} \, \vec \sigma_A \otimes \vec \sigma_B)
\in S.
\eeq
This seems a good $\rho_0$ for $w_\alpha$ if $1/3 < \alpha \leq 1$;
and we use it for $\rho^{\, '}$ in the Theorem part iii), Eq.(\ref{TheoremVar}).
With $\rho_0 - w_\alpha  = \frac{1}{4} (\alpha - 1/3) \,
\vec \sigma_A \otimes \vec \sigma_B$
and $\|\vec \sigma_A \otimes \vec \sigma_B\|_2 = 2 \sqrt{3}$ we get
\beq
D(w_\alpha) \leq \frac{\sqrt{3}}{2} \, (\alpha - 1/3).
\eeq
On the other hand, the observable which according to Eq.(\ref{Amax}) violates the GBI
(\ref{GBI}) maximally is
$A = - \, \vec \sigma_A \otimes \vec \sigma_B/2\sqrt{3}$.
In fact,
\beq
\left( w_\alpha \, \bigg| \, - \frac{\vec \sigma_A \otimes \vec \sigma_B}{2\sqrt{3}}
\right) = \alpha \; \frac{\sqrt{3}}{2}
\eeq
and a pure product $\rho$ gives
$(\rho \, | \, \vec \sigma_A \otimes \vec \sigma_B) = \vec n \cdot \vec m$.
Since $| \vec n \cdot \vec m | \leq 1$ and this cannot be increased by mixing
we have proved
$B(w_\alpha) \geq \frac{\sqrt{3}}{2} (\alpha - 1/3)$.
But $D$ and $B$ can be written as $\min_\rho \max_A$ and
$\max_A \min_\rho$ of $(\rho - w \, | \, A)$ and generally
$\min \max \geq \max \min$ so a priori we know $D(w) \geq B(w)$.
Therefore the above inequalities imply
\beq
D(w_\alpha) = B(w_\alpha) = \frac{\sqrt{3}}{2} \, (\alpha - 1/3) \qquad
\forall \; 1/3 \leq \alpha \leq 1.\\
\eeq
Furthermore the minimizing $\rho_0$ is given by Eq.(\ref{rho0}) and the
maximizing observable is $- \, \vec \sigma_A \otimes \vec \sigma_B/2\sqrt{3}$.
Considering the state with $\alpha = 1$ we finally get
\beq\label{BIsinglet}
(\rho \, | - \vec \sigma_A \otimes \vec \sigma_B ) \leq 1
\quad \forall \; \rho \in S \quad \mbox{and} \quad
(w_{\alpha = 1} | - \vec \sigma_A \otimes \vec \sigma_B) = 3 \, ,
\eeq
and the GBI is violated by a factor 3.
But this ratio is not significant since by $A \ra A + c{\mathbf 1}$ it can
be given any value. Meaningful is $B(w)$ since it is not affected by this
change.

For the parameter values $-1/3 \leq \alpha \leq 1/3$ the states $w_{\alpha}$
(\ref{Wstate}) are separable, for $1/3 < \alpha < 1$ they are mixed entangled,
and the limit $\alpha = 1$ represents the spin singlet state
which is pure and maximally entangled.\\

Let us consider next the tangent functionals. From expression (\ref{Amax}) we get
our old friend, the flip operator \cite{Werner}
\beq
A_t = \frac{1}{4} \, ({\mathbf 1} + \vec \sigma_A \otimes \vec \sigma_B) \, .
\eeq
It is not positive but applying the transposition operator $T$, defined by
$T (\sigma^i)_{kl} = (\sigma^i)_{lk} \,$, on Bob it turns into a positive operator
\beq
({\mathbf 1}_A \otimes T_B) \, A_t = \frac{1}{4} \, ({\mathbf 1} + \sigma_A^x \otimes \sigma_B^x
- \sigma_A^y \otimes \sigma_B^y + \sigma_A^z \otimes \sigma_B^z) \, ,
\eeq
which can be nicely written as $4 \times 4$ matrices\\
\beq
{A_t} = \frac{1}{4}
\left( \ba{cccc}
2 & 0 & 0 & 0\\
0 & 0 & 2 & 0\\
0 & 2 & 0 & 0\\
0 & 0 & 0 & 2\\
\ea \right)
\qquad
{({\mathbf 1}_A \otimes T_B) \, A_t} = \frac{1}{4}
\left( \ba{cccc}
2 & 0 & 0 & 2\\
0 & 0 & 0 & 0\\
0 & 0 & 0 & 0\\
2 & 0 & 0 & 2\\
\ea \right) \, .
\eeq
Operator $A_t$ is not only tangent functional for the mixed separable state
$\rho_0$ (\ref{rho0}) but with
\beqa
( \rho|A_t ) &=& \frac{1}{16} \, \mbox{Tr } \big[ ({\mathbf 1} + n_i \,
\sigma_A^i \otimes {\mathbf 1}_B + m_i \, {\mathbf 1}_A \otimes \sigma_B^i  + n_i m_j \,
\sigma_A^i \otimes \sigma_B^j)
( {\mathbf 1} + \vec \sigma_A \otimes \vec \sigma_B ) \big] \no \\
&=& \frac{1}{4} \, (1 + \vec n \cdot \vec m) = 0
\eeqa
it is a tangent functional for all pure separable states with $\vec m = - \vec n$,
which is especially the case for those states used for $\rho_0$ (\ref{rho0}).
This illustrates point iii) of the properties of the set $S$.

On the other hand, for the pure separable states in this face we
can find other tangent functionals. For example, for the state
\beq\label{rhoz}
\rho_z = \frac{1}{4} \, ({\mathbf 1} + \sigma_A^z \otimes {\mathbf 1}_B +
{\mathbf 1}_A \otimes \sigma_B^z + \sigma_A^z \otimes \sigma_B^z)
\eeq
we easily see within $4 \times 4$ matrices that the operators
\beq
{\rho_z} = \frac{1}{4}
\left( \ba{cccc}
1 & 0 & 0 & 0\\
0 & 0 & 0 & 0\\
0 & 0 & 0 & 0\\
0 & 0 & 0 & 0\\
\ea \right)
\eeq
and
\beq\label{A-t}
{A_t} = \frac{1}{a^2 + b^2}
\left( \ba{cccc}
0 & 0 & 0 & ab\\
0 & a^2 & 0 & 0\\
0 & 0 & b^2 & 0\\
ab & 0 & 0 & 0\\
\ea \right)
\qquad
{({\mathbf 1}_A \otimes T_B) \, A_t} = \frac{1}{a^2 + b^2}
\left( \ba{cccc}
0 & 0 & 0 & 0\\
0 & a^2 & ab & 0\\
0 & ab & b^2 & 0\\
0 & 0 & 0 & 0\\
\ea \right) \, > \, 0
\eeq
satisfy the requirement of a tangent functional. For the state $\rho_x$
(let $z \ra x$ in Eq.(\ref{rhoz})), however, we have
$(\rho_x|A_t) = 1$. \\

\noindent
{\bf Remark:}

\noindent
At this stage we would like to compare our approach to generalized Bell inequalities
with the more familiar type of inequalities
(compare also with Refs. \cite{TerhalPhysLett,WernerWolf}).
Usually the BI is given by an operator in the tensor product, where
by classical arguments only some range of expectation values can be expected,
whereas the quantum case permits an other range. In our case, classically we would
expect
\beq
0 \leq (\rho_{class}|{\mathbf 1} + \vec \sigma_A \otimes \vec \sigma_B) \leq 2
\quad \mbox{or} \quad |(\rho_{class}|\vec \sigma_A \otimes \vec \sigma_B)| \leq 1
\eeq
because the expectation value of the individual spin is maximally $1$ and the largest
(smallest) value should be obtained when they are parallel (antiparallel).
This range of expectation values can exactly be achieved by all
separable states $\rho \in S$, whereas we can find an entangled quantum state,
the spin singlet state $w_{\alpha = 1}$ (\ref{Wstate}), which gives
\beq
(w_{\alpha = 1}|{\mathbf 1} + \vec \sigma_A \otimes \vec \sigma_B) = - 2
\quad \mbox{or} \quad |(w_{\alpha = 1}|\vec \sigma_A \otimes \vec \sigma_B)| = 3 \, .
\eeq
This demonstrates that the tensor product operator
$\vec \sigma_A \otimes \vec \sigma_B$ cannot be written as a CHSH operator,
where the ratio is limited by $\sqrt 2$. If we perturb a pure
separable state like
\beq
\rho_{\ve} = \frac{1}{4} \,\big({\mathbf 1} + n_i \, \sigma_A^i \otimes {\mathbf 1}_B
- n_i \, {\mathbf 1}_A \otimes \sigma_B^i  - (n_i n_j + \ve_{ij}) \,
\sigma_A^i \otimes \sigma_B^j \big)
\eeq
then the expectation value
\beq
(\rho_{\ve}|{\mathbf 1} + \vec \sigma_A \otimes \vec \sigma_B) = \cO (\ve)
\eeq
is of order $\cO (\ve)$, as the operator constructed in Ref.\cite{Gisin},
which shows the sensitivity of $\A_t$ (\ref{A-t}) as entanglement witness.\\

In the familiar Bell inequality derived by CHSH \cite{CHSH}
\beq\label{CHSH-BI}
(\rho | A_{CHSH}) \leq 2 \, ,
\eeq
with $\rho \in S$ (actually CHSH consider classical states $\rho_{class}$,
a generalization of separable states,
in their work \cite{CHSH}), a rather general observable (a 4 parameter family
of observables)
\beq\label{A_CHSH}
A_{CHSH} \,\, = \,\, \vec a \cdot \vec \sigma_A \otimes (\vec b - \vec b^{\, '})
\cdot \vec \sigma_B \,\, + \,\, \vec a^{\, '} \cdot \vec \sigma_A \otimes
(\vec b + \vec b^{\, '}) \cdot \vec \sigma_B
\eeq
is used, where $\vec a, \vec a^{\, '}, \vec b, \vec b^{\, '}$ are any unit vectors
in ${\mathbf R}^3$.

\noindent
However, the spin singlet state $w_{\alpha = 1}$ (\ref{Wstate}) gives
\beq
(w_{\alpha = 1} | A_{CHSH}) \, = \, - \, \vec a \cdot (\vec b - \vec b^{\, '}) \,
- \, \vec a^{\, '} \cdot (\vec b + \vec b^{\, '}) \, ,
\eeq
which violates the CHSH inequality (\ref{CHSH-BI}) maximally
\beq
(w_{\alpha = 1} | A_{CHSH}) = 2 \, \sqrt{2} \, ,
\eeq
for appropriate angles: $(\vec a \, ,\vec b) = (\vec a^{\, '},\vec b) =
(\vec a^{\, '},\vec b^{\, '}) = 135^o, \, (\vec a \, ,\vec b^{\, '}) = 45^o$,
whereas in this case we find (for all separable states $\rho \in S$)
\beq
\max_{\rho \,\in S} \, (\rho | A_{CHSH}) = \sqrt{2} \, .
\eeq

Bell in his original work \cite{Bell64} considers only 3 different directions
in space (which corresponds to the specific case
$\vec a^{\, '} =  - \, \vec b^{\, '}$ in CHSH (\ref{A_CHSH}))
and assumes a strict anti-correlation
\beq
(\, \rho \, | \, \vec a^{\, '} \cdot \vec \sigma_A \, \otimes \, \vec a^{\, '}
\cdot \vec \sigma_B \, ) \, = \, - \, 1 \, .
\eeq
Then he derives the inequality
\beq\label{Bell-BI}
(\rho | A_{Bell}) \leq 1 \, ,
\eeq
(which clearly follows from (\ref{CHSH-BI}) under the mentioned conditions),
where now the observable is
\beq\label{A_Bell}
A_{Bell} \,\, = \,\,  \vec a \cdot \vec \sigma_A \otimes (\vec b - \vec b^{\, '})
\cdot \vec \sigma_B \,\, - \,\, \vec b^{\, '} \cdot \vec \sigma_A \otimes
\vec b \cdot \vec \sigma_B \, .
\eeq
The expectation value of Bell's observable in the spin singlet state
\beq
(w_{\alpha = 1} | A_{Bell}) \, = \, - \, \vec a \cdot (\vec b - \vec b^{\, '}) \,
+ \, \vec b^{\, '} \cdot \vec b
\eeq
lies (maximally) outside the range of BI (\ref{Bell-BI})
\beq
(w_{\alpha = 1} | A_{Bell}) = \frac{3}{2} \, ,
\eeq
for the angles $(\vec a \, ,\vec b^{\, '}) = (\vec b^{\, '},\vec b) = 60^o, \,
(\vec a \, ,\vec b) = 120^o$, whereas now we have for all anti-correlated separable
states $\rho_a = \{ \rho \in S \, | \, \mbox{with} \, \vec n \cdot \vec m = - 1 \}$
\beq\label{A-Bellmax}
\max_{\rho_a \,\in S} \, (\rho | A_{Bell}) = \frac{3}{4} \, .
\eeq
Note that generally $\forall \rho \in S$ the maximum (\ref{A-Bellmax}) is larger,
namely $\sqrt{3}/2$ instead of $3/4$.\\

We observe that the maximal violation of the GBI, Eq. (\ref{BImax}), is largest for
our observable $- \, \vec \sigma_A \otimes \vec \sigma_B$ ,
where the difference between singlet state and separable state is $2$
(recall Eq.(\ref{BIsinglet})), whereas in case of CHSH it is $\sqrt{2}$
and in Bell's original case it is $3/4 \,$.

Although the violation of BI's is a manifestation of entanglement, as a criterion
for separability it is rather poor. There exists a class of
entangled states which satisfy the considered BI's, CHSH (\ref{CHSH-BI}),
Bell (\ref{Bell-BI}) but not our GBI (\ref{BI}) or (\ref{BIsinglet}).
For a given entangled state there exists always some operator (entanglement
witness) so that it satisfies the GBI for separable states but not for this
entangled state. The class of these operators can be obtained by the positivity
condition of Ref.\cite{HHH96}. On the other hand, as criterion for nonlocality
the violation of the familiar BI's is of great importance.\\

Let us finally return again to the geometry of the quantum states
(see also Ref. \cite{VerstraeteDehaeneMoor}).
For two spins there is a one parameter family of equivalence
classes of pure states, interpolating between the separable one and the one
containing $w_{\alpha = 1}$. The latter is quite big and contains 4 orthogonal
projections, the ``Bell states''. They are obtained by rotating $\vec \sigma_A$ by
$180^o$ around each of the axis:
\beqan
w_{\alpha = 1}
&=& \frac{1}{4}({\mathbf 1} - \sigma_A^x \otimes \sigma_B^x - \sigma_A^y \otimes \sigma_B^y
- \sigma_A^z \otimes \sigma_B^z) =: P_0 \\
&\ra& \frac{1}{4}({\mathbf 1} - \sigma_A^x \otimes \sigma_B^x + \sigma_A^y \otimes \sigma_B^y
+ \sigma_A^z \otimes \sigma_B^z) =: P_1 \\
&\ra& \frac{1}{4}({\mathbf 1} + \sigma_A^x \otimes \sigma_B^x - \sigma_A^y \otimes \sigma_B^y
+ \sigma_A^z \otimes \sigma_B^z) =: P_2 \\
&\ra& \frac{1}{4}({\mathbf 1} + \sigma_A^x \otimes \sigma_B^x + \sigma_A^y \otimes \sigma_B^y
- \sigma_A^z \otimes \sigma_B^z) =: P_3 \, .
\eeqan
However, there are far more since $\sigma_A$ and $\sigma_B$ can be rotated
independently.\\

\begin{figure}
\centering
\includegraphics[width=6cm,height=6cm,keepaspectratio=true]{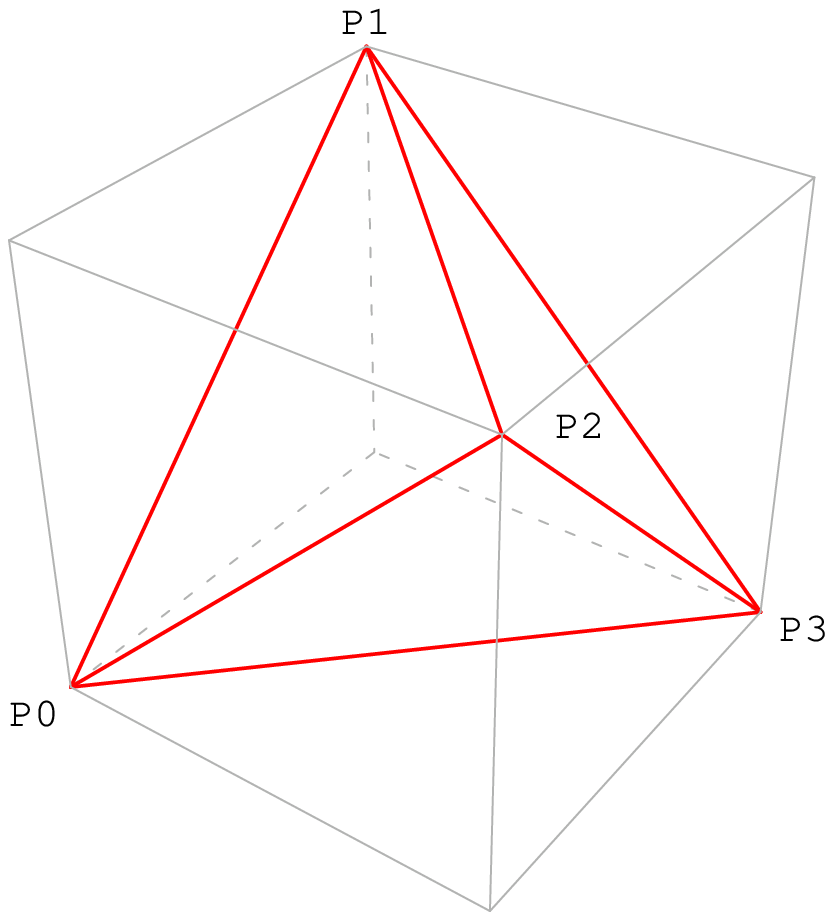}
\includegraphics[width=6cm,height=6cm,keepaspectratio=true]{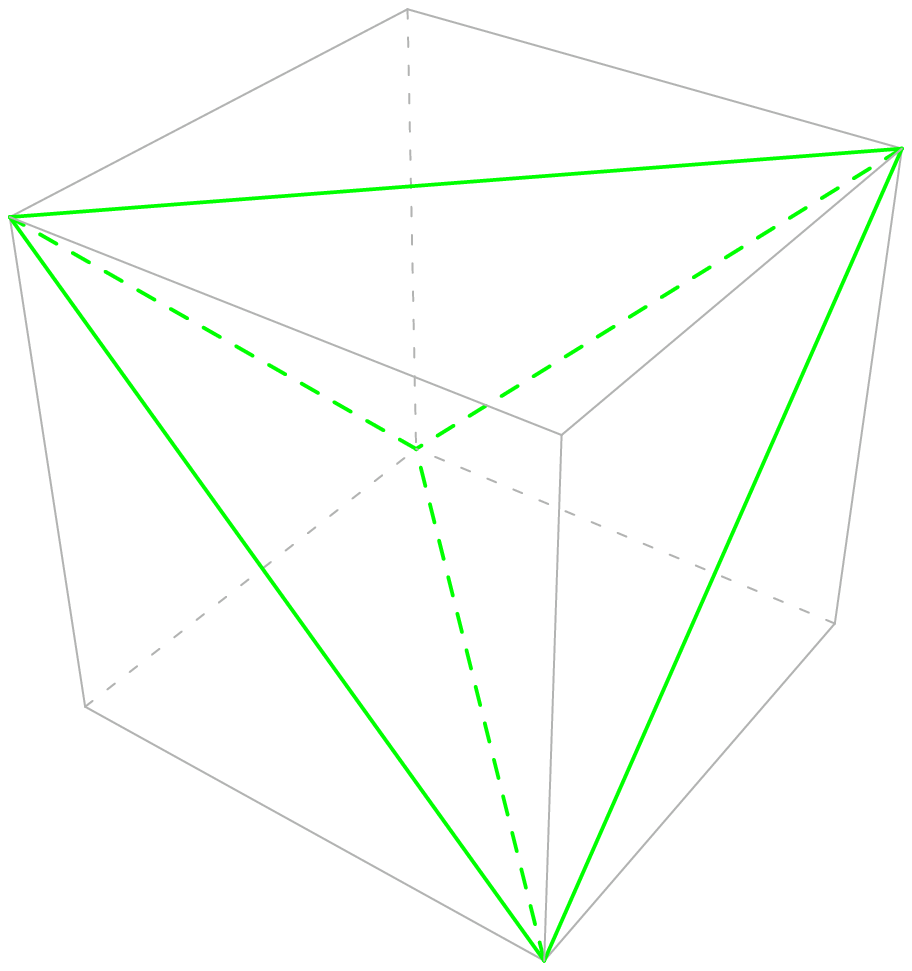}
\includegraphics[width=6cm,height=6cm,keepaspectratio=true]{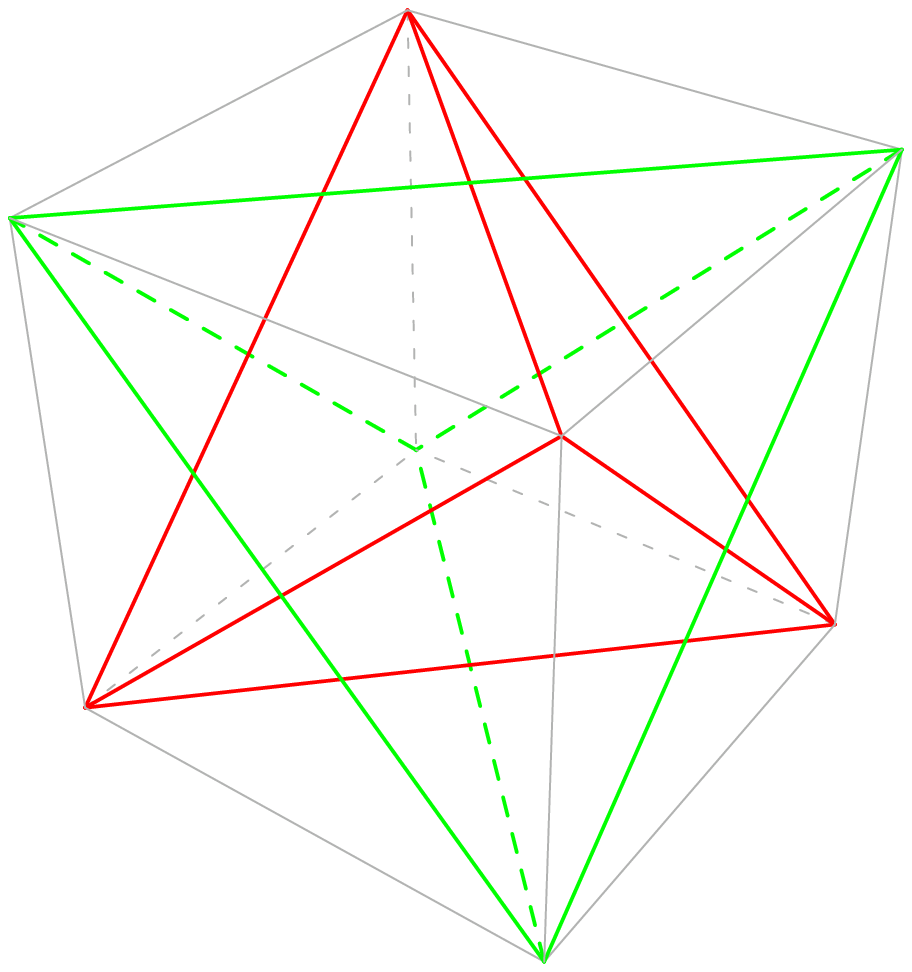}
\includegraphics[width=6cm,height=6cm,keepaspectratio=true]{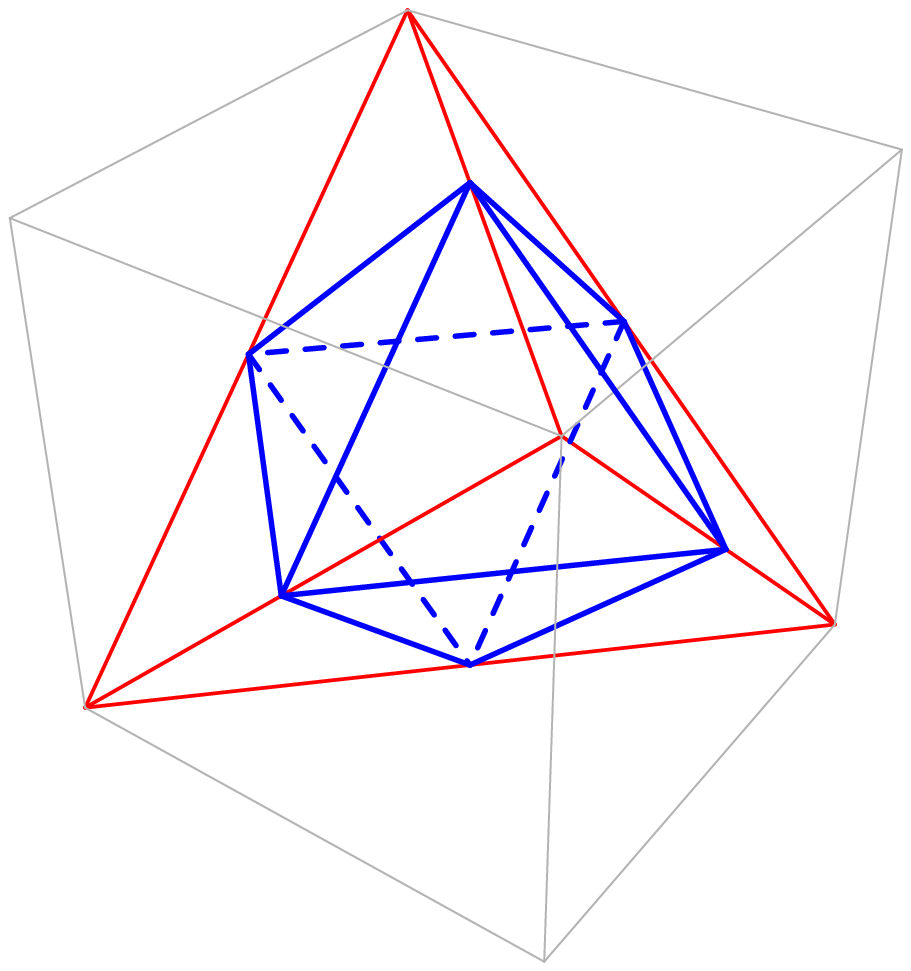}
\caption{In the left figure above we have plotted the tetrahedron of states
described by the density matrix $w_c$ (\ref{w-c}) in the $\vec c$-space
and to the right the reflected set of states $({\mathbf 1}_A \otimes T_B) \, w_c$
is shown. In the left figure below we have plotted the intersection of the two sets
($w_c$ and its mirror image) and, finally, to the right
the double-pyramid of separable states $S \cap \{ w_c \} \,$.}
\end{figure}

\vspace{0.1cm}

The matrix $c_{ij}$ in Eq.(\ref{observable}) will in general not be diagonalizible
but by two independent orthogonal transformations on both spins it can be
diagonalized. Thus the correlation part of a density matrix $w_c$ contains
3 parameters $c_i$:
\beq\label{w-c}
w_c = \frac{1}{4} \Big({\mathbf 1} + \sum_{i=1}^3 c_i \; \sigma_A^i \otimes
\sigma_B^i \Big).
\eeq
Density matrix $w_c$ can be expressed as convex combination of the projectors onto the 4
Bell states. Positivity requires that the $c_i$ are contained in the
convex region spanned by the 4 points $(-1,-1,-1)$, $(-1,1,1)$,
$(1,-1,1)$, $(1,1,-1)$. This region is screwed and the intersection with
its mirror image -- compare with point vi) of the properties of $S$ --
characterizes the separable states
$\sum_{i=1}^3 |c_i| \leq 1$. Reflection in $c$-space is effected by time
reversal on one spin and not on the other (``partial transposition'') and
the classically correlated states form the set invariant under this
transformation. These properties are illustrated in Fig. 3 (see also
Refs. \cite{HH,VollbrechtWerner}).\\

Finally, we would like to mention that the quantum states which
are used in the model for decoherence of entangled systems in
particle physics \cite{BGH,BG} also lie in the regions of the plotted
separable and entangled states.\\

\section{Summary and conclusion}

In this article we have used tangent functionals on the set of separable states
as entanglement witnesses defining a generalized Bell inequality. The operators
are vectors in the Hilbert space $\Ha_s$ with Hilbert Schmidt norm. We show that
the euclidean distance of an entangled state to the separable states is equal to
the maximal violation of the GBI  with the tangent functional as entanglement witness.
This description gives a nice geometric picture of separable and entangled states
and their boundary, especially in the example of two spins. The advantage of considering
the larger set of GBI's is that they are a criterion for separability (or entanglement)
whereas the usual BI's are not.

\vspace{1cm}

\noindent
{\bf Acknowledgements:}

We are thankful to Katharina Durstberger for her drawings and to Fabio Benatti,
\u Caslav Brukner, Franz Embacher, Walter Grimus, Beatrix Hiesmayr
and Anton Zeilinger for fruitful discussions.
We also thank Frank Verstraete and Jens Eisert for useful comments.
The research was performed within the FWF Project No. P14143-PHY of the
Austrian Science Foundation.

\end{document}